\documentclass[prb,showpacs,floatfix,twocolumn]{revtex4}

\usepackage{graphicx}

\begin{document}

\title[CoSe$_2$O$_5$: A Low Dimensional Antiferromagnet]
{Magnetic Structure and Susceptibility of CoSe$_2$O$_5$: A Low Dimensional 
Antiferromagnet} 

\author{Brent C. Melot, Brian Paden, and Ram Seshadri}
\affiliation{Materials Department and Materials Research Laboratory\\
	 University of California, Santa Barbara CA 93106}

\author{Ambesh Dixit and Gavin Lawes}
\affiliation{Department of Physics and Astronomy, Wayne State University\\
Detroit, MI 48201}

\author{Emmanuelle Suard}
\affiliation{Institute Laue Langevin, F-38042 Grenoble, France}

\begin{abstract}

CoSe$_2$O$_5$ has a crystal structure consisting of zig-zag chains of edge 
shared CoO$_6$ octahedra running along the $c$ axis, with the chains separated 
by Se$_2$O$_5^{2-}$ units. Magnetic susceptibility measurements indicate 
a transition at 8.5\,K to an ordered state. We investigate here the nature 
of this magnetic ordering using magnetization and specific heat measurements in 
addition to powder neuttron diffraction. A transition to long-range antiferromagnetic 
order is found below $T_N$ = 8.5\,K as identified by magnetic susceptibility
measurements and magnetic Bragg reflections, with a propagation vector 
$\mathbf{k}$ = 0. The magnetic structure shows that the moments align 
perpendicular to the $c$-axis, but with the spins canting with respect to the 
$a$ axis by, alternately, $+$8$^\circ$ and $-$8$^\circ$. Interestingly, the 
low-field magnetic susceptibility does not show the anticipated cusp-like 
behavior expected for a well-ordered antiferromagnet. When the susceptibility 
is acquired under field-cooling conditions under a 10\,kOe field, the 
the usual downturn expected for antiferromagnetic ordering is obtained.
Heat capacity measurements at low temperatures indicate the presence of 
gapped behavior with a gap of 6.5\,K.

\end{abstract}


\maketitle

\section{Introduction}

Recently there has been a significant interest in materials with reduced 
crystallographic dimensionality because of their strong interplay between 
charge, lattice, and magnetic degrees of 
freedom.\cite{Cheong2007,Kimura2007} From the perspective of magnetism, 
one-dimensional chains of spins are especially interesting because of the 
possibility of competing nearest and next-nearest neighbor exchange 
interactions.\cite{Park2007} Examples of recently studied compounds with 
magnetic chains include ones showing geometric frustration, magnetoelectric
coupling, and possibly quantum tunneling of the magnetization, as exemplified
by Ca$_3$Co$_2$O$_6$,\cite{Hardy2003,Maignan2004,Choi2008} 
MnWO$_4$,\cite{Taniguchi2006} and LiCu$_2$O$_2$.\cite{Masuda2005,Park2007}

Here we report on CoSe$_2$O$_5$, an orthorhombic compound with the $Pbcn$
space group, consisting of chains of edge sharing CoO$_6$ octahedra which 
zig-zag along the the $c$ axis. Each chain is bound together by 
Se$_2$O$_5^{2-}$ units \textit{via} shared oxygen at the corners of the 
octahedra. The diselenite units can be visualized as two trigonal pyramids of 
SeO$_3$ which share a corner and contain two lone-pairs of electrons which 
point in antiparallel directions. This connectivity packs the chains in 
hexagonal arrays and isolates them magnetically. This compound and its 
structure was reported by Harrison \textit{et al.}\cite{Harrison1992}
but no reports on the properties have been made. We were particularly
interested in this compound because of its low dimensionality, the presence of
orbitally degenerate octahedral Co$^{2+}$, and the presence of possibly
symmetry reducing lone pairs.

We have used a combination of magnetic susceptibilty and magnetization 
measurements, powder neutron diffraction, and specific heat measurements to 
characterize the nature of the magnetic ground state below 8.5\,K and 
at low magnetic fields. We also suggest the presence of more complex ground 
states at high fields, a subject for future study. The compound displays a 
spin gap, with the size of the gap being  6.5\,K.

\section{Experimental methods}

CoSe$_2$O$_5$ was prepared following the reported hydrothermal 
procedure.\cite{Harrison1992} SeO$_2$ (5.0\,g, Cerac, 99.99\%) was dissolved 
in 15\,cm$^3$ of water and combined with CoSO$_{4}\cdot$xH$_2$O 
(2.0\,g, Sigma-Aldrich, 98\%). The mixture was sealed in a 23-mL 
poly(tetrafluoroethylene)-lined pressure vessel (Parr Instruments) and heated 
to 200\,$^\circ$C for 48 hours. The resulting product consisted of dark purple 
single crystals averaging 1.5\,mm\,$\times$\,1.5\,mm\,$\times$\,0.5\,mm.

ZnSe$_2$O$_5$, a non-magnetic analogue to the title compound, was also prepared 
to give an estimate of the lattice contribution to the specific heat.
SeO$_2$ (0.7317\,g, Cerac, 99.99\%) was ground in an agate mortar with ZnO 
(0.2683\,g, Sigma-Aldrich, 98\%), sealed in a silica tube, and heated at 
350$^\circ$C for 48 hrs.\cite{Meunier1974zso} 
The resulting product consisted of an off-white polycrystalline powder.

Neutron diffraction data were collected on a sample of well-ground single 
crystals at the D2B powder diffractometer at the Institut Laue-Lagnevin (ILL), 
France~\cite{D2B} using a wavelength of 1.5943\,\AA\/.
The wavelength of the incident neutrons was determined by refining the data 
obtained at 300\,K and fixing the cell parameters to the values determined 
from a room temperature x-ray diffraction pattern collected on a Philips XPERT 
MPD diffractometer operated at 45\,kV and 40\,mA.
In order to achieve a better fit to the lowest lying magnetic reflections, a 
diffraction pattern was also collected at 2\,K using a wavelength of 
2.399\,\AA\/. 

Temperature dependence of the DC magnetization were measured on well-ground 
powder samples using a Quantum Design MPMS 5XL SQUID magnetometer. Powders
were preferred since the shape and dimensions of the crystal made it dificult 
to align a single crystal with respect to the field direction.
The specific heat data was collected on a 9.5\,mg single crystal using the
semiadiabatic technique as implemented in a Quantum Design Physical Property 
Measurement System (PPMS), under zero applied field, as well as as under a
50\,kOe field. The measurement on non-magnetic ZnSe$_2$O$_5$ was made by
mixing the compound with equal parts by mass of Ag powder and pressing into a 
pellet in order to improve thermal coupling. The contribution from Ag
was measured separately and subtracted.

\section{Results and Discussion} 

\begin{table*}[t]
\caption{Summary of results from the Rietveld structure refinement of the variable temperature neutron diffraction patterns}
\label{tab:refine}
\begin{tabular}{lrrrrrrrrr}
\hline \hline
  & 300\,K  & 150\,K  & 20\,K & 15\,K & 10\,K  & 5\,K & 2\,K & 2.4 (\AA)\\
\hline
$a$ (\AA)         & 6.7911   & 6.7890 (1)  & 6.7900 (2)  & 6.7897 (2)  & 6.7892 (2)   & 6.7897  (2) & 6.7893 (2)  & 6.78877 (2)  \\
$b$ (\AA)         & 10.366   & 10.353 (2)  & 10.3524 (2) & 10.3522 (2) & 10.3516 (2)  & 10.3529 (1) & 10.3523 (1) & 10.35166 (4) \\
$c$ (\AA)         & 6.0750   & 6.0580 (2)  & 6.0515 (1)  & 6.0514 (1)  & 6.0512 (1)   & 6.0527 (2)  & 6.0524 (1)  & 6.05195 (2)  \\
$M$ ($\mu_B$)     &          &          &             &             &              & 2.80        & 2.84        & 3.04 \\
$V$ (\AA$^3$)     & 427.66   & 425.80(2)& 425.38(2)   & 425.34(2)   & 425.28(2)    & 425.47(2)   & 425.39(2)   & 425.31(3) \\
R$_{Bragg}$ (\%)  & 2.94     & 2.38     & 2.35        & 2.39        & 2.43         & 2.34        & 2.43        & 2.39 \\
R$_{magn}$ (\%)   &          &          &             &             &              & 5.06        & 5.31        & 4.76 \\
$\chi^2$          & 2.87     & 2.78     & 2.92        & 2.81        & 2.90         & 3.11        & 3.11        & 1.60  \\
\hline \hline
\end{tabular}
\end{table*}

\begin{figure}[t]
\centering \includegraphics[width=7cm]{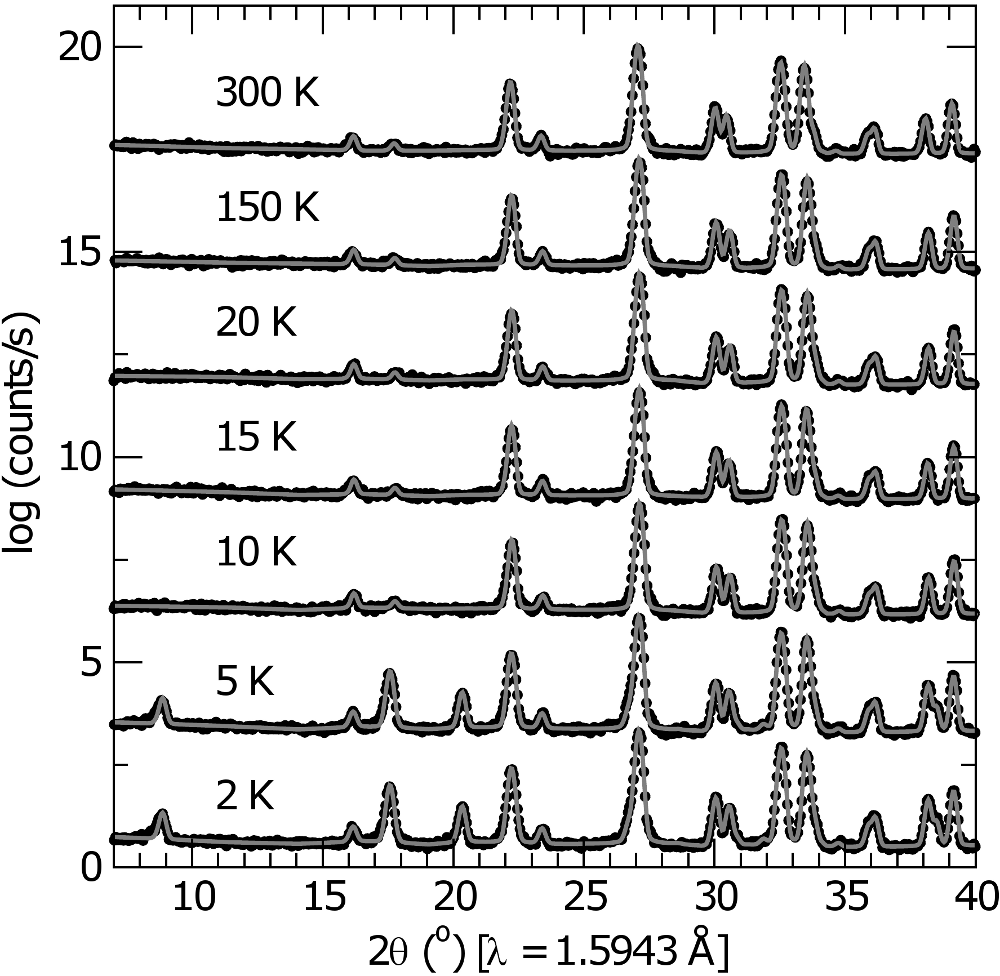}
\caption{Low angle region of the powder neutron-diffraction patterns of 
CoSe$_2$O$_5$ (D2B, ILL). Note the development of magnetic Bragg peaks 
in the 5\,K and 2\,K patterns.}
\label{fig:neutron}
\end{figure}

\begin{figure}[t]
\centering \includegraphics[width=7cm]{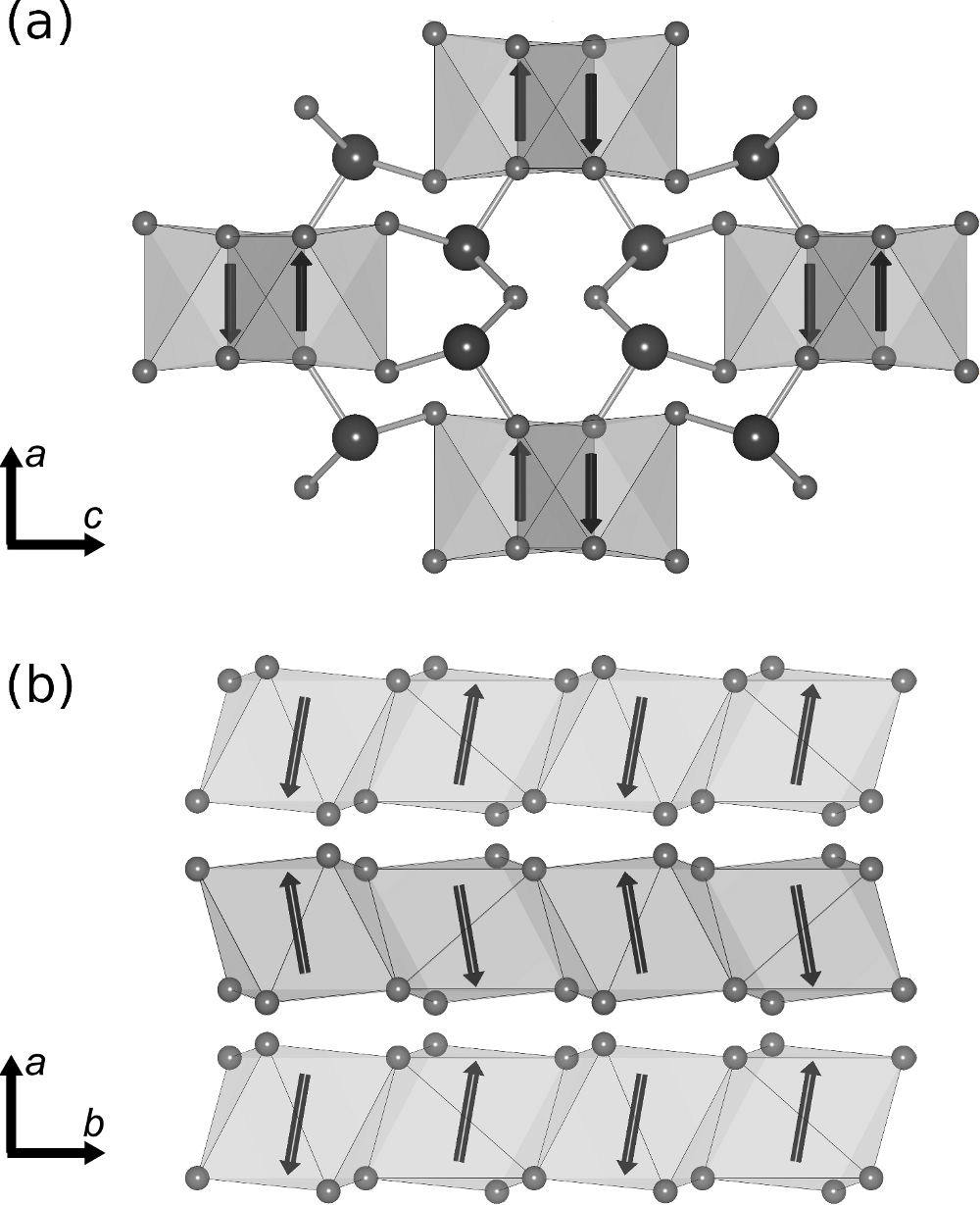}
\caption{Magnetic structure of CoSe$_2$O$_5$ as determined from Rietveld 
refinements of the neutron diffraction pattern obtained using a 
$\lambda$ = 2.4\,\AA\/ at 2\,K. (a) View down the $b$-axis of the $Pbcn$ 
structure. (b) View down the $c$-axis. The small light grey atoms are oxygen
while the larger and darker grey atoms are selenium}
\label{fig:magstruct}
\end{figure}

\begin{figure}[ht]
\centering \includegraphics[width=7cm]{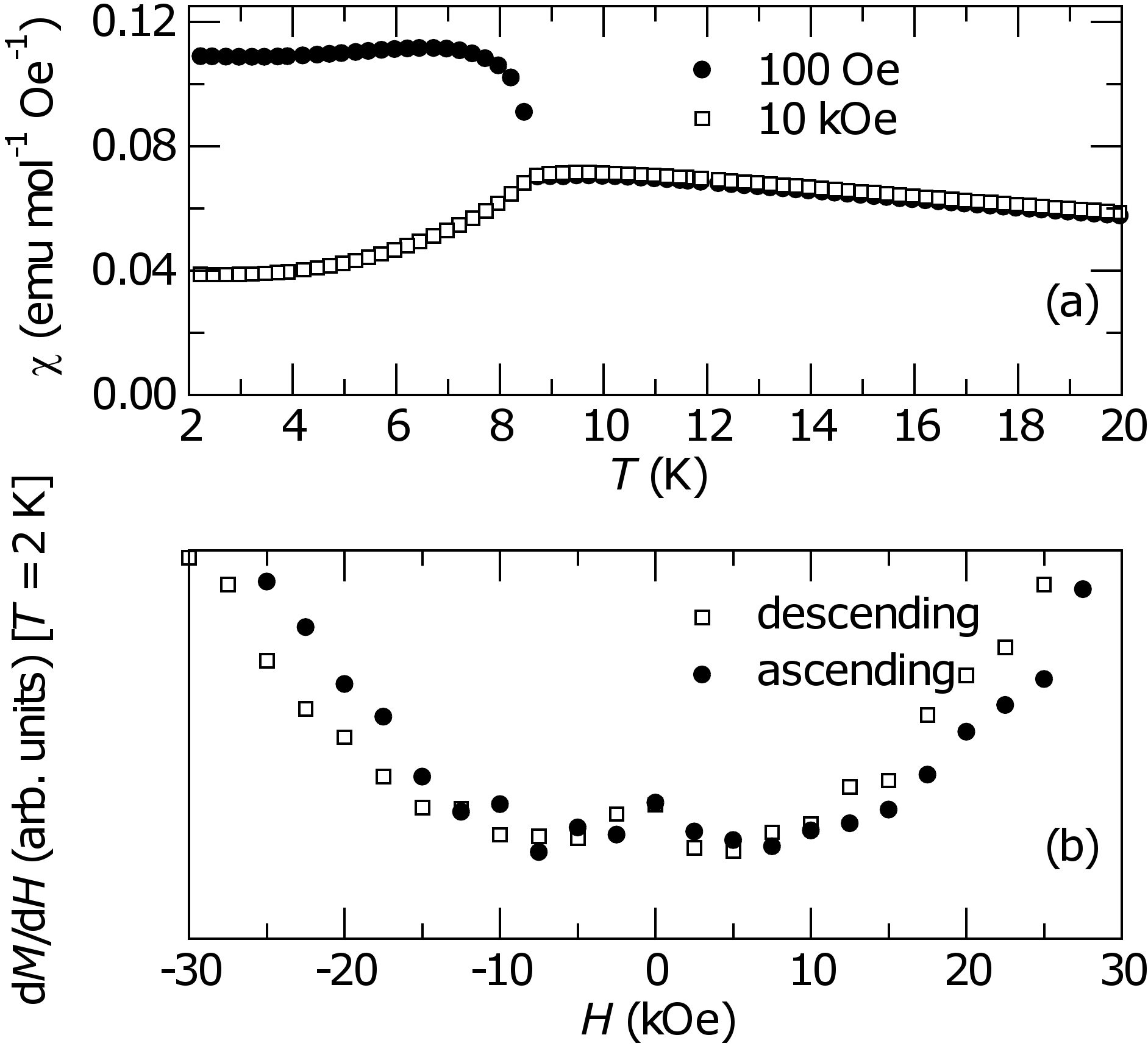}
\caption{(a) Magnetic susceptibility a powder sample of CoSe$_2$O$_5$ acquired
after field cooling under fields of 100\,Oe and 10\,kOe. Note the existence
of a weak ferromagnetic component in the low-field data, which does not appear
under the higher magnetic fields. Plot of $\mbox{d}M/\mbox{d}{H}$ at 
$T$ = 2\,K showing the existence of a field-induced magnetic transition 
for fields near $H$ = 5\,kOe.}
\label{fig:magn}
\end{figure}

The nuclear and magnetic structure of CoSe$_2$O$_5$ were refined using the 
Rietveld method as implemented in the \textsc{fullprof} software 
suite.\cite{FullProf} The peak shape was described using the 
Thompson-Cox-Hastings pseudo-Voigt function, and the background was fit by 
interpolation between regions showing no Bragg reflections. 
The crystal structures were analysed based on the model 
proposed by Harrison \textit{et al.}\cite{Harrison1992}, although it should be 
noted that we use the standard setting of space group number 60 ($Pbcn$) 
instead of the alternate $Pnab$ description originally used. A summary of key 
refinement results is shown in table\,\ref{tab:refine} and in 
table\,\ref{tab:struct}. For the magnetic structure, a group 
theoretical analysis was performed using representational analysis as 
implemented in the program \textsc{sara{\it h}}\cite{SARAH} was used to 
determine all of the possible spin configurations which were compatible with 
the crystal symmetry. 

The thermal evolution of the neutron-diffraction patterns collected from 
300\,K to 2\,K are shown in fig. \ref{fig:neutron}.
Three magnetic reflections appear below 10\,K at 8$^\circ$, 17$^\circ$, and 
20$^\circ$. These three peaks could only be fit simultaneously by using the 
basis vectors of the irreducible representation $\Gamma_8$.  The resulting 
magnetic structure associated with $\Gamma_8$ and refined from the 2.4\,\AA\/ 
data is illustrated in fig. \ref{fig:magstruct}.  
Each chain of Co moments align antiferromagnetically down the length of the 
chain and with respect to the neighboring chains. 
The moments cant in the $ac$ plane, forming, alternately, angles of 
$+$8$^\circ$ and $-$8$^\circ$ with respect to the $a$ axis. The moments have
no component along $b$. The magnetic moment on every Co atoms refined to a 
value of 3\,$\mu_B$ at 2\,K using the $\lambda$ = 2.399\,\AA\/ neutron 
diffraction data. This corresponds to three unpaired spins per Co atom in the
ordered antiferromagnetic structure.

The high temperature region (350\,K to 400\,K) of the inverse susceptibility 
 was fit to the Curie-Weiss equation, $C/(T-\Theta_{CW})$, to 
obtain the effective moment $\mu_{eff}$ from the Curie constant, and the 
Curie-Weiss intercept $\Theta_{CW}$. A Curie-Weiss temperature of $-$34\,K and 
an effective moment of 5.19\,$\mu_B$ were extracted from the fits to the data 
collected under a 100\,Oe field. This value for the effective is in close 
agreement for the completely unquenched and decoupled orbital contribution 
($L+S$) of 5.2\,$\mu_B$ expected from octahedrally coordinated Co$^{2+}$ 
($d^3$,$t_{2g}^{5}e_g^2$). This effective moment also agrees well with the 
three unpaired electrons refined in the neutron data, confirming that the 
system is in a high spin state. The frustration index, 
$f = |\Theta_{CW}|/ T_N$, is close to 4 which indicates a moderate degree of 
frustration in this structure.

\begin{table*}[ht]
\caption{Atomic positions and anisotropic thermal parameters of 
CoSe$_2$O$_5$ at 300 and 2\,K}
\label{tab:struct}
\begin{tabular}{lrrr|rrrrrr}
\hline \hline
  & $x$  & $y$  & $z$ & $\beta_{11}$  & $\beta_{22}$ & $\beta_{33}$ & $\beta_{12}$ & $\beta_{13}$ & $\beta_{23}$\\
\hline
\multicolumn{10}{c}{300\,K} \\
\hline
Co     & 0      & 0.940  & 0.75 & 0.0024 & 0.0001 & 0.0020 &         & -0.0020 &         \\
Se     & 0.631  & 0.154  & 0.968 & 0.0013 & 0.0007 & 0.0051 &  0.0003 & -0.0007 &  0.0002 \\
O1     & 0.805  & 0.210  & 0.342 & 0.0027 & 0.0012 & 0.0098 &  0.0003 &  0.0004 & -0.0001 \\
O2     & 0.5  & 0.932  & 0.250 & 0.0063 & 0.0010 & 0.0109 &         & -0.0054 &         \\
O3     & 0.840  & 0.933  & 0.452 & 0.0007 & 0.0020 & 0.0067 & -0.0015 & -0.0017 &  0.0004 \\
\hline
\multicolumn{10}{c}{2\,K} \\
\hline
Co     & 0      & 0.937  & 0.75 & 0.0127 & 0.0042 & 0.0128 &         & -0.0014 &         \\
Se     & 0.630  & 0.154  & 0.969 & 0.0099 & 0.0043 & 0.0148 &  0.0002 &  0.0002 &  0.0000 \\
O1     & 0.807  & 0.211  & 0.342 & 0.0109 & 0.0047 & 0.0168 &  0.0000 &  0.0004 & -0.0004 \\
O2     & 0.5  & 0.931  & 0.250 & 0.0112 & 0.0043 & 0.0167 &         & -0.0003 &         \\
O3     & 0.840  & 0.933  & 0.451 & 0.0099 & 0.0050 & 0.0167 & -0.0004 &  0.0000 &  0.0000 \\
\hline \hline
\end{tabular}
\end{table*}

Fig.\,\ref{fig:magn}(a) shows the temperature dependence of the magnetic 
susceptibility in small and large external magnetic fields for a powder sample
of CoSe$_2$O$_5$. When cooled in small fields, a sharp jump in the 
susceptibility is observed below 8.5\,K. This behavior is unusual and
is not in keeping with the magnetic structure which suggests complete 
antiferromagnetic ordering of spins. It is proposed that the unusual
transition in the susceptibility might be associated with the anisotropic,
one-dimensional nature of the magnetic structure, and also perhaps, with the
magnetism being centered on ooctahedral Co$^{2+}$ which is orbitally degenerate.
Data acquired after field cooling in the larger field of 10\,kOe results in 
the more usual downturn in the susceptibility expected from an exact 
cancellation of all moments. 

To more carefully probe the dependence of the magnetization on magnetic field 
in the ordered phase, we measured $M(H)$ curves at $T$ = 2\,K. The low field 
behaviour can be seen most clearly by plotting the differential magnetization 
($\mbox{d}M/\mbox{d}H$), as shown in fig.\ref{fig:magn}(b), showing a small 
maximum near $H$ = 0, with broad minima near $H$ = 5\,kOe, with no significant 
dependence on the direction of the magnetic field sweep. Recall that the canted 
spin structure observed in the neutron scattering experiment was not subject to 
any external field.  These magnetization measurements suggest that even a small 
external magnetic field can perturb the spin structure towards a more collinear 
arrangement. Further neutron scattering studies under a magnetic field are 
planned to explore the nature of magnetic ordering as a function of both field 
and temperature. 

\begin{figure}[t]
\centering \includegraphics[width=7cm]{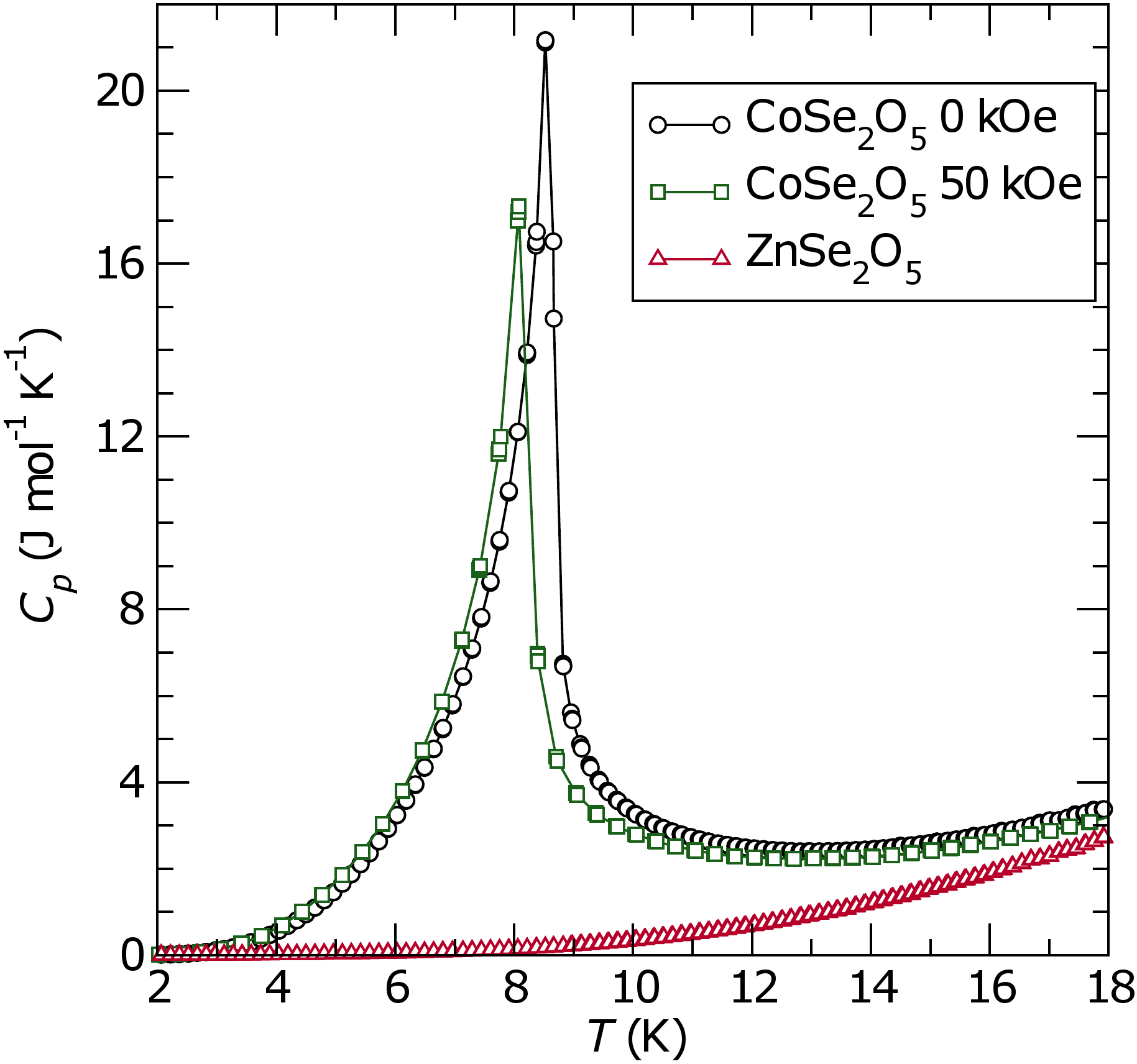}
\caption{Temperature dependence of the specific heat of CoSe$_2$O$_4$
measured on a 9.5\,mg single crystal under zero field, and under a 
$H$ = 50\,kOe field. A nonmagnetic analogue, ZnSe$_2$O$_5$, was also measured 
in order to obtain the lattice contribution to the specific heat.
Note the large width of the ordering peak and the release of entropy well 
above the transition temperature of 8.5\,K.}
\label{fig:hc}
\end{figure}

The temperature dependence of the specific heat $C_p$ of CoSe$_2$O$_5$ shown in 
Fig.\ref{fig:hc} exhibits a lamda-type anomaly around 8.5\,K corresponding to 
the transition to long-range magnetic order observed in the susceptibility and 
in the neutron diffraction data. Measurements on non-magnetic ZnSe$_2$O$_5$ 
give an estimate of the lattice contribution using a three term expansion with 
the best fit yielding a Debye temperature, $\Theta_D,$ of 222\,K. 
An estimate of the change in entropy associated with the magnetic transition 
can be obtained by integrating $C_{p,mag}/T$ defined between the specific heat
data of CoSe$_2$O$_5$ and ZnSe$_2$O$_5$. The change in entropy due to the 
magnetic transition, thus determined, was 5.31\,J\,mol$^{-1}$\,K$^{-1}$ which 
is much smaller than the theoretical value of 11.53\,J\,mol$^{-1}$\,K$^{-1}$ 
predicted by the Boltzmann equation ($\Delta S = R\ln(2S+1)$, $S=3/2$). 
The difference between expected and measured magnetic entropy changes could
arise from the canted nature of the antiferromagnetic ground state.
Such diminished entropy changes at magnetic transitions are a common feature of 
frustrated magnetic materials.

Additionally the large width of the ordering peak at 8\,K with a width close
to 15\,K as compared to simple non-frustrated antiferromagnets, which 
typically are only a few K wide\cite{Skalyo1964}) also reflects the release of 
entropy at temperatures higher than the ordering temperature, which is a common 
feature of geometrically frustrated systems. A slight downward shift of the 
heat capacity peak was found when a 50\,kOe field was applied, but there was 
little change in the magnitude or shape of the peak.

\begin{figure}[t]
\centering \includegraphics[width=7cm]{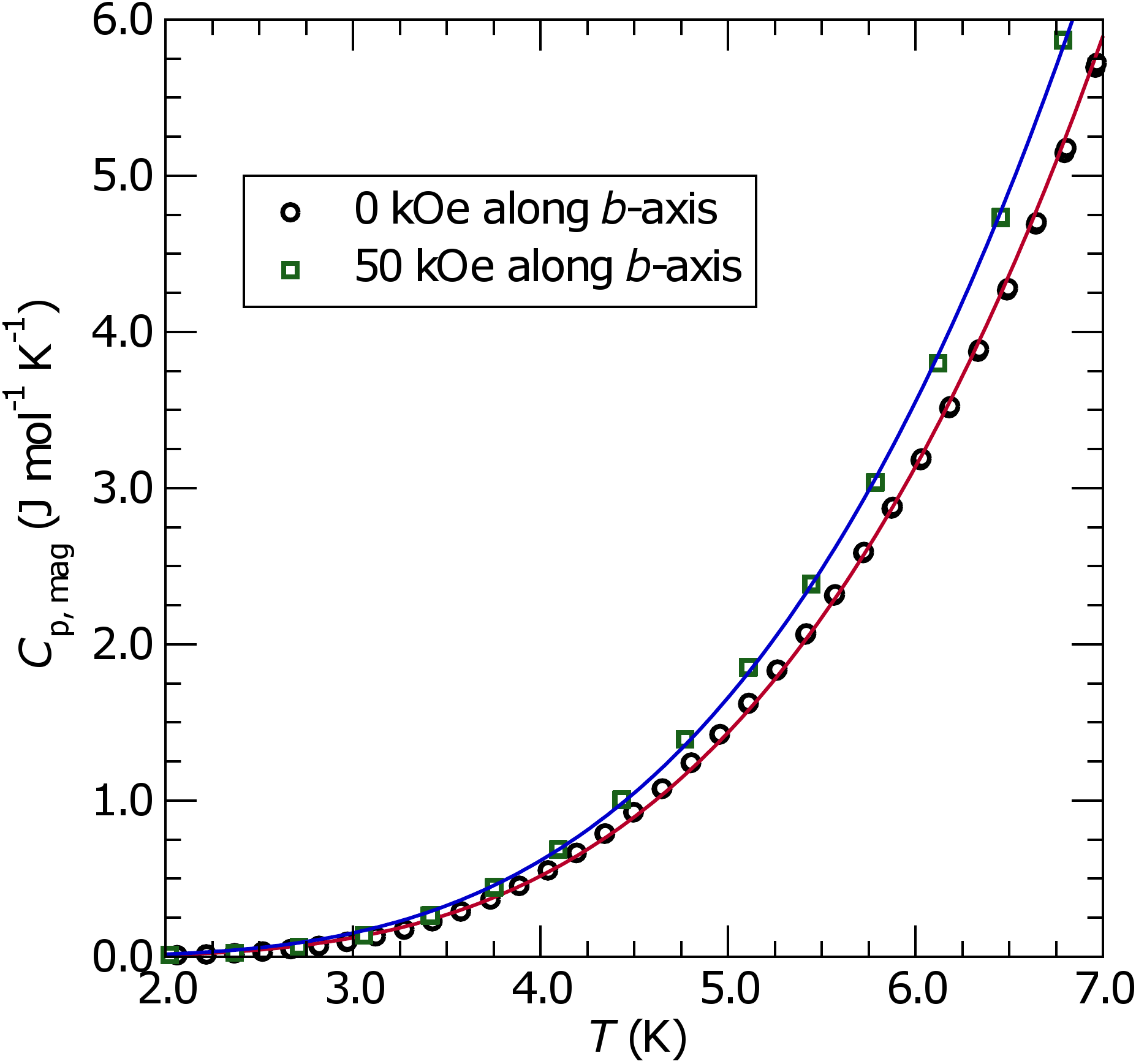}
\caption{Magnetic specific heat of CoSe$_2$O$_4$ below the ordering temperature
fitted to the formula for a gapped spin system, $C_p$ $\propto$ 
$T^3\exp(-\Delta/k_BT)$, with $\Delta$ = 6.5\,K for data acquired under 
$H$ = 0\,kOe as well as $H$ = 50\,kOe.}
\label{fig:hcfit}
\end{figure}

A fit to the specific heat below the transition temperature could not be 
obtained using only the $T^3$ term as would be expected for an 
antiferromagnet. Instead, the best fit to the data was found when 
an exponential term was included, using 
$C_{p,mag} \propto T^3\exp(-\Delta / k_BT)$.\cite{TariBook} Such an analysis 
reflects the presence of low-lying magnetic excitations with an energy 
gap $\Delta = 6.5 K$. The presence of spin gaps in one-dimensional magnetic 
materials is well known.\cite{Mikeska1991} 

\subsection{Conclusions}

We have examined the magnetic properties of CoSe$_2$O$_5$ using neutron 
diffraction, magnetization and magnetic susceptibility, and specific heat
measurements. We show that below the magnetic ordering temperature, the one 
dimensional magnetic chains arrange their moments antiferromagnetically down 
the length of the chain and with respect to neighboring chains.
The magnetic order in small fields is found to exhibit some canting, although 
the net moment is zero. Under large applied field, the magnetic behavior
changes, pointing to a rich $H-T$ phase diagram in this system.
We see evidence for low lying magnetic excitations with a spin gap of 
6.5\,K, that are also possibly implicated in the field-induced phase transition
seen in $\mbox{d}M/\mbox{d}H$ \textit{vs.} $H$ at 2\,K.

\section*{Acknowledgements} We gratefully acknowledge the National Science 
Foundation for support through Career Awards to RS (DMR 0449354) and 
to GL (DMR 06044823) and for the use of MRSEC facilities at UCSB
(DMR 0520415).

\clearpage

\end{document}